\documentclass[reprint,amsmath,amssymb,aps,prb]{revtex4-2}

\usepackage{graphicx}
\usepackage{dcolumn}
\usepackage{bm}
\usepackage{xcolor}
\makeatletter
\renewcommand{\@seccntformat}[1]{}
\makeatother

\begin{document}

\preprint{APS/123-QED}

\title{Emergent Superfluidity of Hard-Core Excitons in Single-Layer Breathing-Kagome Nb$_3$Te$_x$Cl$_{8-x}$}

\author{Mahtab A. Khan$^{1,2}$ and Michael N. Leuenberger$^{1,3}$}
\email{michael.leuenberger@ucf.edu}

\affiliation{$^1$NanoScience Technology Center, University of Central Florida, Orlando, FL 32826, USA}
\affiliation{$^2$Department of Physics, Federal Urdu University of Arts, Sciences and Technology, Islamabad 44000, Pakistan}
\affiliation{$^3$Department of Physics and College of Optics and Photonics (CREOL),
University of Central Florida, Orlando, FL 32816, USA}
\date{\today}%




\date{\today}

\begin{abstract}
We develop a microscopic theory of superfluidity for hard-core dark excitons on the triangular lattice by mapping the large-$U$ Bose--Hubbard model to an effective XXZ spin-$\tfrac{1}{2}$ Hamiltonian including virtual hopping processes. Within this framework, we identify the superfluid phase that emerges between the two Mott-insulating endpoints at fillings 0 and 1, and derive its mean-field structure via a canted-spin solution. We then construct the corresponding continuum Landau-Ginzburg (LG) functional and analyze phase fluctuations and vortex dynamics. In two dimensions, the superfluid--normal transition is shown to be governed by a Berezinskii--Kosterlitz--Thouless (BKT) mechanism with a stiffness determined by microscopic parameters. Our results provide a unified description connecting lattice-scale exciton dynamics to continuum critical behavior in triangular geometries.
\end{abstract}

\maketitle


\paragraph{Introduction---}The interplay between strong correlations, lattice geometry, and topology continues to drive the search for novel collective phases in single-layer (SL) quantum materials. Excitonic systems provide a particularly promising platform, as their charge neutrality allows coherent bosonic motion even in the presence of sizable Coulomb interactions. In SL breathing kagome lattices such as Nb$_3$Cl$_8$~\cite{khan2024multiferroic}, dark excitons originating from $d_{z^{2}}$ orbitals form a robust hard-core bosonic manifold at low densities, yet their collective phases on frustrated lattices remain largely unexplored. In this work, we investigate the emergence of superfluidity for such hard-core excitons on the triangular lattice by deriving an effective XXZ Hamiltonian from the large-$U$ Bose--Hubbard model including virtual doublon processes. This microscopic mapping enables us to connect lattice-scale exciton dynamics to continuum critical behavior, revealing a Berezinskii--Kosterlitz--Thouless (BKT) transition governed by the superfluid stiffness determined directly from the underlying microscopic parameters.

Exciton superfluidity in low dimensions has long been a central goal in condensed-matter physics, with the earliest compelling evidence emerging in bilayer quantum Hall systems, where counterflow transport and interlayer coherence demonstrated the formation of exciton condensates under extreme magnetic fields and ultralow temperatures\cite{eisenstein2004bose}. More recently, van der Waals heterostructures have provided a versatile platform for exploring excitonic many-body states in zero magnetic field. In particular, experiments on WSe$_2$/MoSe$_2$ bilayers have revealed signatures of interlayer exciton coherence and superfluid crossover \cite{ma2021strongly,cutshall2025imaging}, supported by the moiré-induced spatial confinement of excitons. Despite these advances, the microscopic origin of superfluidity in strongly interacting \emph{dark} excitons subject to hard-core constraints, especially on intrinsically frustrated triangular geometries, remains largely unexplored. Our work addresses this gap by establishing a controlled Bose--Hubbard to XXZ mapping and connecting it to a continuum BKT framework.

\begin{figure*}[!t]\includegraphics[width=\textwidth]{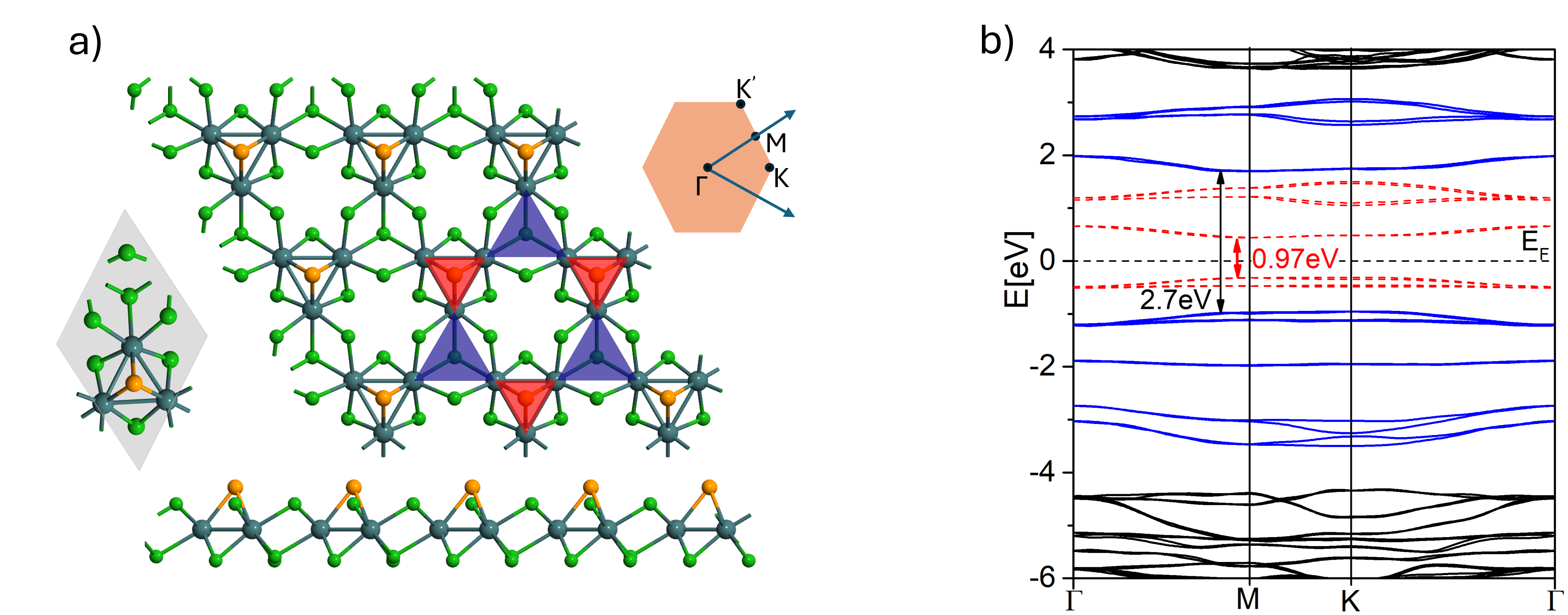}
\centering
\caption{a) Top and side view of Nb$_3$TeCl$_7$, shaded grey (orange) region (rhombus) is the unit cell (Brillouin zone) of Nb$_3$TeCl$_7$ consisting of 3 Nb (dark green balls) 1 Te (orange balls) and 7 Cl (light green balls) atoms. The breathing kagome lattice is built from Nb atoms forming irregular hexagons enclosed by two inequivalent sets of equilateral triangles with different areas (shaded red and blue). SL Nb$_3$TeCl$_7$ has C$_{3v}$ symmetry. b)  GW band structure of SL Nb$_3$TeCl$_7$, showing substantial increase in the electronic band gap E$_g^{\textrm{GW}}=$2.7 eV as compared with the PBE band gap of E$_g^{\textrm{PBE}}=$0.97 eV. Blue (red) colored lines show flat bands obtained through GW (PBE) -calculations. Black solid lines correspond to the continuum of states with dispersion.}
\label{fig:Nb3Cl8_Structure_Bs}
\end{figure*}

While bilayer structures have so far provided the most direct experimental access to excitonic superfluidity—whether in quantum Hall bilayers or in recent WSe$_2$/MoSe$_2$ heterostructures—the requirement of a double-layer geometry limits the range of accessible lattice phenomena. In contrast, dark excitons in SL breathing kagome lattices such as Nb$_3$Cl$_8$, offer a fundamentally different setting in which the excitonic manifold itself forms a correlated lattice system. Previous work\cite{khan2024multiferroic} has shown that breathing kagome single-layers (SLs) naturally host an exciton Mott insulating ground state, formed by strong on-site interactions that localize one dark exciton per trimer site. Bulk Nb$_3$Cl$_8$ does not host a bound exciton below the electronic band gap because the strong three-dimensional dielectric screening suppresses the electron–hole attraction such that the exciton binding energy is insufficient to overcome the quasiparticle gap. In contrast, the reduced dielectric screening and enhanced Coulomb interactions in the SL Nb$_3$Cl$_8$, stabilize a tightly bound dark exciton in the ground state. This localized excitonic state forms the basis of the exciton Mott insulating phase at unit filling. In this limit, dark excitons in Nb$_3$Cl$_8$ may well be described within the hard core Bose-Hubbard model.
However for such a system, presence of vacancies provide mobile empty sites into which neighboring excitons can hop, continuously tuning the system away from unit filling may enable a transition from the immobile Mott state to a coherent exciton superfluid at fractional filling.

Based on the experimental synthesis of Nb$_3$TeCl$_7$ in Ref.~\onlinecite{zhang2023topological}, our work is based on the fact that Te substitution in Nb$_3$Cl$_8$, forming Nb$_3$Te$_{x}$Cl$_{8-x}$ with variable Te concentration $x$, creates intrinsic vacancies in the excitonic manifold by removing the lowest-energy dark exciton on selected trimer sites. These vacancies provide empty sites that enable exciton hopping, thereby generating a hard-core bosonic degree of freedom on an intrinsically frustrated triangular network. We further evaluate the BKT transition temperature and find that it reaches a maximum at the particle-hole-symmetric point $n=1/2$, i.e. at half filling of the hard-core exciton lattice. This mechanism yields a purely single-layer realization of exciton superfluidity driven by vacancy-induced quantum dynamics in a strongly interacting excitonic lattice. Our results open a pathway toward identifying experimental signatures of 
vacancy-driven exciton superfluidity in SL Nb$_3$Cl$_8$ derivatives, including vortex-mediated transport anomalies and characteristic optical responses near the BKT transition. \textcolor{black}{In addition, exciton superfluidity provides a charge-neutral transport channel, strongly reducing impurity scattering and Joule heating. As a result, excitons can carry spin or valley information with greatly enhanced coherence, offering a natural pathway toward low-power spintronics.}



\textcolor{black}{\paragraph{Fractional-Filling Superfluidity---}Dark Frenkel excitons in Nb$_3$Cl$_8$ are tightly localized on a single trimer, so placing two excitons on the same site forces their electron and hole wavefunctions to overlap within the same molecular orbitals. This overlap induces strong Pauli exclusion and large Coulomb penalties, yielding an on-site repulsion $U$ on the order of hundreds of meV to eV, similar to the electronic case in bulk Nb$_3$Cl$_8$ \cite{Grytsiuk2024}---far exceeding the inter-trimer hopping $t$. Double occupancy is therefore effectively forbidden, and in the following we employ the hard-core Bose--Hubbard description.}

\textcolor{black}{In the hard-core limit, the excitonic manifold reduces to a two-level local Hilbert space, enabling the standard mapping of the hard core Bose--Hubbard model to an XXZ spin-$\tfrac12$ Hamiltonian i.e.
\begin{equation}
H_{\mathrm{XXZ}} 
= - J_{xy} \sum_{\langle ij \rangle}
\left( S_i^{+} S_j^{-} + S_i^{-} S_j^{+} \right)
+ J_z \sum_{\langle ij \rangle} S_i^{z} S_j^{z},
\label{eq:XXZ}
\end{equation}
with $J_{xy}= t$ and $J_{z}=-4t^{2}/U$ generated by virtual doublon processes. This mapping naturally relates the exciton density to the spin canting angle via $n=\tfrac12(1+\cos\theta)$. Here $n$ denotes
the exciton filling per trimer, while $x=1-n$ represents the corresponding vacancy
(hole) concentration. So that $n=0$ and $n=1$ correspond to empty $\theta=\pi$ (Nb$_3$TeCl$_7$) and
fully occupied dark spin triplet exciton states $\theta=0$ (Nb$_3$Cl$_8$), respectively, while any fractional filling $0<n<1$ corresponds to
a canted configuration with $0<\theta<\pi$. Within this framework, any non-integer (fractional) filling corresponds to a canted spin configuration with a finite transverse component, implying a stable superfluid phase. The transverse order---and hence the superfluid stiffness---is maximized near excitonic half filling ($n=1/2$) where $\theta=\pi/2$ \cite{matsubara1956lattice, matsuda1957lattice}. In terms of exciton density, these well-established results provide the microscopic foundation for the vacancy-induced superfluidity explored in this work. Detailed derivations are provided in SI(S1).}

\textcolor{black}{Although mean-field theory suggests a finite order parameter at fractional filling, true long-range order cannot exist in two dimensions at finite temperature because of strong phase fluctuations. Nevertheless, the system still undergoes a BKT transition, which depends only on the phase stiffness---rather than the mean-field order parameter---and therefore remains well defined in 2D for fractional filling.}
 
\paragraph{Electronic and Excitonic Structure of Nb$_3$TeCl$_7$---}Our model system consists of a single layer (SL) of Nb$_3$TeCl$_7$, 
where a layer of Nb atoms is sandwiched between layers of Cl/Te atoms, held together through strong covalent bonds as shown in Fig.~\ref{fig:Nb3Cl8_Structure_Bs} (a). The breathing kagome lattice is built from Nb atoms forming irregular hexagons enclosed by two inequivalent sets of adjacent equilateral triangles with different areas (shaded red and blue), as shown in Fig.~\ref{fig:Nb3Cl8_Structure_Bs} (a). The electronic band structure of Nb$_3$Cl$_8$ is largely determined by the symmetry properties of the high-symmetry points and lines in the first Brillouin zone of its triangular lattice.  Fig.~\ref{fig:Nb3Cl8_Structure_Bs} also illustrates both the unit cell and the Brillouin zone of Nb$_3$Cl$_8$. The crystal structure of single-layer (SL) Nb$_3$Cl$_8$ exhibits C$_{3v}$ symmetry. 

We consider a unit cell of SL Nb$_3$Cl$_8$, consisting of 11 atoms in a hexagonal structure, with edge lengths $a=b=6.81$ \AA. A vacuum layer of $20$~\AA~is introduced in the $z$-direction to minimize interactions between the periodic images of Nb$_3$Cl$_8$.
\begin{figure*}[!t]\includegraphics[width=\textwidth]{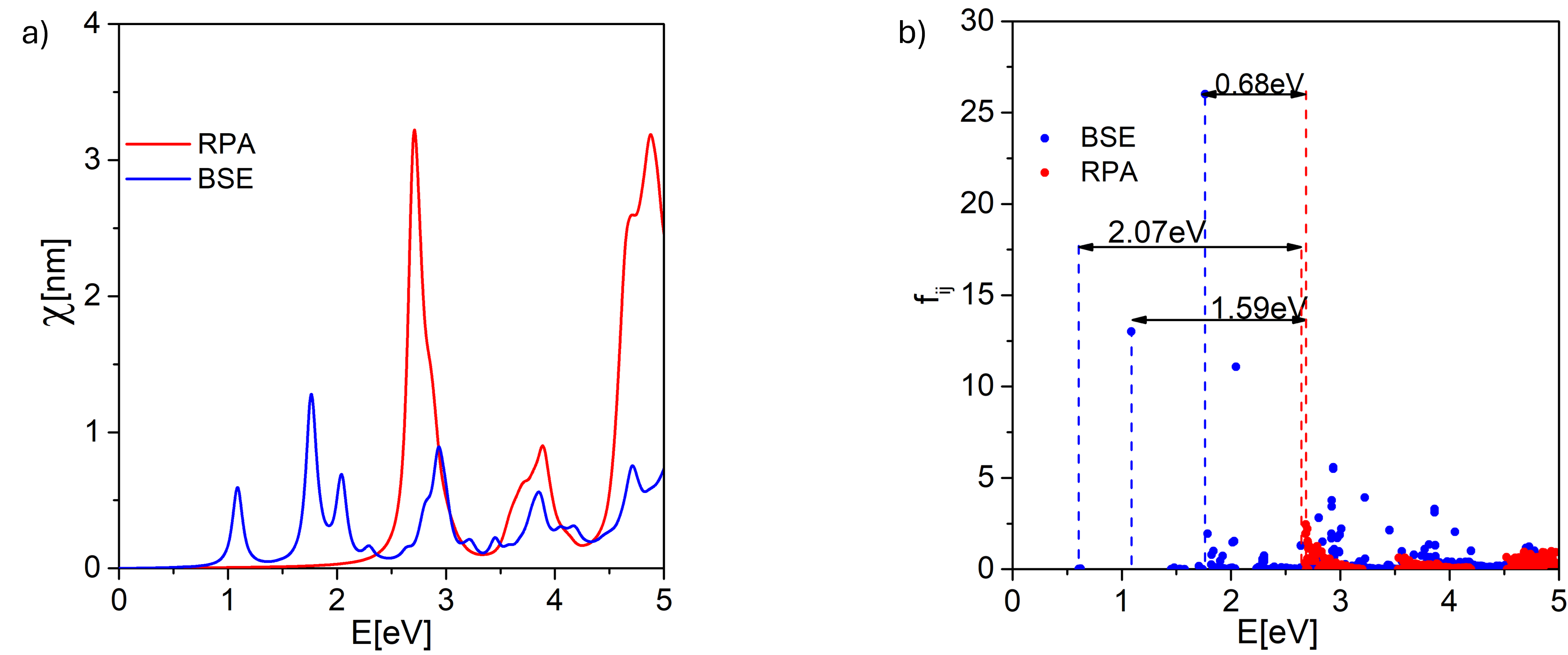}
\centering
\caption{a) The absorption spectra of SL Nb$_3$TeCl$_7$, shown without (red curve) and with (blue curve) electron-hole interactions, exhibit a prominent peaks at 1.075 eV and 1.75~eV. b) The oscillator strength f$_{ij}$ as a function of energy reveals key insights into the system's excitonic properties. The first two bright exciton peaks can be seen at 1.075 eV and 1.75 eV with binding energies 1.59 eV and 0.68 eV, respectively.}
\label{fig:susceptibility_TD}
\end{figure*}

We first perform mean field DFT in generalized gradient approximation (GGA) as implemented in VASP,\cite{vasp} on a 9$\times$9$\times$1 k-grid with a cut-off energy of 400 eV. For PBE Calculations a band gap of 0.97 eV is obtained Fig.~\ref{fig:Nb3Cl8_Structure_Bs} (b) (dashed line). We use the PBE non-collinear SOC results as the starting point for our GW calculations.

The GW calculations were initiated using DFT-derived wavefunctions. Self-consistency was achieved by iteratively updating the GW eigen-energies, thereby removing dependence on the single-particle energies obtained from the initial DFT computations.
For GW calculations we consider a 9$\times$9$\times$1 grid with a cut-off energy (ENCUT) of 400 eV. For response function the we set ENCUTGW$=$200 eV. The parameter ENCUTGW controls the basis set for the response functions in exactly the same manner as ENCUT for the wave functions. We use 324 empty bands,

For BSE calculations, we consider 20 bands above and below the Fermi level, it is important to note that results seem to converge at 12 bands above and below the Fermi level.

Fig.~\ref{fig:Nb3Cl8_Structure_Bs} (b) presents the band structure results from noncollinear SOC calculations, incorporating self-energy corrections through GW calculations. Nearly flat bands with a very small dispersions can be seen, well separated from the continuum of bands and are therefore available for direct experimental observation.  A substantial increase of 1.73 eV in the band gap with the GW correction compared to the PBE (SI) highlights the importance of self-energy corrections in Nb$_3$Cl$_8$. It is interesting to note that flat bands appear in the form of triplets, i.e. a singlet accompanied by a two-fold degenerate doublet. The existence of triplets is a consequence of three-fold rotational symmetry of the crystal~\cite{Erementchouk_MoS2,Khan_TMDCs}.

Fig.~\ref{fig:susceptibility_TD} presents the absorption spectra with oscillator strengths for SL Nb$_3$Cl$_8$, both with and without electron-hole interactions. The first two bright excitons appear at energies of 1.075~eV and 1.75~eV, with binding energies of 1.59~eV and 0.68~eV, respectively, as shown in Fig.~\ref{fig:susceptibility_TD}. These unusually high binding energies reflect the weak dielectric screening in SL Nb$_3$TeCl$_7$, a characteristic feature of many two-dimensional materials.\cite{MoS2_EBE,Luo_2025} Such strong excitonic effects highlight the potential of SL Nb$_3$TeCl$_7$ for optoelectronic applications, particularly in devices that rely on strong light–matter interactions.



\paragraph{Absence of a Bound Excitonic Ground State in Nb$_3$TeCl$_7$---} A key distinction between Nb$_3$Cl$_8$~\cite{khan2024multiferroic} and Nb$_3$TeCl$_7$ is revealed in their BSE excitation spectra. In Nb$_3$Cl$_8$, the lowest excitonic solution lies \emph{below} the RPA onset and appears at a negative energy relative to the quasiparticle continuum. This dark state therefore constitutes a genuine many-body ground-state configuration and reflects the presence of a naturally stabilized, trimer-centered exciton on each Nb$_3$ unit. In contrast, Nb$_3$TeCl$_7$ shows no excitonic state below the continuum: its lowest BSE solution occurs at a positive excitation energy and carries finite oscillator strength. Hence, excitons in Nb$_3$TeCl$_7$ must be created optically, while Nb$_3$Cl$_8$ intrinsically hosts localized excitonic configurations as part of its electronic ground state. The presence of some Nb$_3$TeCl$_7$ into the Nb$_3$Cl$_8$ host lattice introduces vacant exciton trimer sites where dark excitons are absent, enabling excitons on neighboring Nb$_3$Cl$_8$, units to delocalize and thereby establishing a superfluid state.

The absence of a negative-energy exciton in Nb$_3$TeCl$_7$ can be understood from the fundamental reorganization of the trimer electronic states. In Nb$_3$Cl$_8$, the spin-polarized Nb$_3$ trimer hosts a partially filled orbital whose electron--hole configuration is strongly localized on the same trimer. The resulting exchange stabilization and large real-space overlap between the band-edge states push the corresponding exciton below the quasiparticle continuum, yielding a dark excitonic ground state. In Nb$_3$TeCl$_7$, however, the trimer becomes non-magnetic and the band-edge states acquire a different symmetry and Nb--Te hybridization pattern. This reduces both the intra-trimer overlap and the exchange-driven binding energy. As a result, no excitonic configuration is able to collapse below the continuum, and all excitonic states remain at positive excitation energy, consistent with the absence of a negative-energy exciton in 
Nb$_3$TeCl$_7$.

\paragraph{LG–BKT Theory---}To describe the exciton superfluid phase emerging from the vacancy-induced hard-core boson manifold, we adopt a standard coarse-grained Landau--Ginzburg description in terms of a complex order parameter $\psi(\mathbf{r}) \sim \langle b(\mathbf{r})\rangle \sim \langle S^+(\mathbf{r})\rangle$. Symmetry considerations constrain the static free-energy functional in two dimensions to the form
\begin{equation}
\label{eq:LG}
F[\psi] = \int d^2 r \left[
\alpha\, |\psi(\mathbf{r})|^2
+ \frac{\beta}{2}\, |\psi(\mathbf{r})|^4
+ K\, |\boldsymbol{\nabla} \psi(\mathbf{r})|^2
\right],
\end{equation}
with $\alpha$ changing sign at the transition, $\beta>0$ ensuring stability, and $K>0$ encoding the stiffness associated with spatial variations of the condensate. Minimizing Eq.~(\ref{eq:LG}) for a uniform state, $\psi(\mathbf{r}) = \psi_0$, yields the standard mean-field solution $|\psi_0|^2 = -\alpha/\beta$ in the superfluid phase ($\alpha<0$) and
$\psi_0 = 0$ in the normal/Mott regime ($\alpha>0$).

Deep in the superfluid phase, amplitude fluctuations are gapped and the
low-energy physics is governed by phase fluctuations. Writing
\begin{equation}
\psi(\mathbf{r}) = \sqrt{|\psi_0|^2 + \delta n(\mathbf{r})}\,
e^{i\theta(\mathbf{r})}
\approx \sqrt{\rho_s}\, e^{i\theta(\mathbf{r})},
\end{equation}
with $\rho_s \equiv |\psi_0|^2$ the condensate (superfluid) density and
neglecting $\delta n$ to leading order, the gradient term in
Eq.~(\ref{eq:LG}) reduces to
\begin{equation}
|\boldsymbol{\nabla}\psi|^2
\simeq \rho_s\, (\boldsymbol{\nabla}\theta)^2,
\end{equation}
and the free energy takes the phase-only XY form
\begin{equation}
\label{eq:phase_only}
F_\theta = \frac{\rho_s^{\mathrm{eff}}}{2}
\int d^2 r\, (\boldsymbol{\nabla}\theta)^2, 
\qquad
\rho_s^{\mathrm{eff}} = 2 K \rho_s,
\end{equation}
where $\rho_s^{\mathrm{eff}}$ is the superfluid stiffness. Within the XXZ mapping, $\rho_s$ is set by the transverse spin component,
$\rho_s \propto \langle S^+ \rangle^2 =S^2 \sin^2\theta$, which for hard-core bosons yields the characteristic dependence
$\rho_s \propto n(1-n)$ on the average exciton density $n$.

The phase-only functional~(\ref{eq:phase_only}) supports topological vortex configurations with integer winding of $\theta$. In two dimensions, the superfluid--normal transition is of BKT type, controlled by the unbinding of vortex--antivortex pairs. The critical temperature is determined by the universal jump condition for the stiffness~\cite{PhysRevLett.39.1201},
\begin{equation}
k_B T_{\mathrm{BKT}} = \frac{\pi}{2}\,\rho_s^{\mathrm{eff}}(T_{\mathrm{BKT}}),
\end{equation}
In the underlying electronic Hubbard model for Nb$_3$Cl$_8$, the relevant kinetic scale is set by the inter-trimer hopping $t$, which Wannier analysis yields as $t =22$\,meV for the lowest trimer molecular orbital \cite{grytsiuk2024nb3cl8}. This inter-trimer hopping controls the dispersion of the lowest exciton band, so the effective exciton hopping amplitude $t$ in the hard-core Bose–Hubbard/XXZ description is naturally of the same order. So that, the stiffness (S2:SM),
\begin{equation}
 \rho_{s}^{(0)}(n)= A\, t\, n(1-n),   
\end{equation} 
shows that $T_{\mathrm{BKT}}$ is finite for any fractional filling $0<n<1$ and is maximized at half filling $n=1/2$, where the transverse spin component---and hence the exciton condensate stiffness---is largest. The factor $A$ is the lattice coarse-graining coefficient that relates the discrete XXZ bond stiffness to the continuum superfluid stiffness.  It arises from the gradient expansion that maps the lattice phase differences to the continuum term $(\nabla \theta)^2$.  For the triangular lattice this procedure yields (S2:SM)
\begin{equation}
A = \frac{z}{\sqrt{3}} = 2\sqrt{3},
\end{equation}
where $z=6$ is the coordination number.  

It is essential that the stiffness entering Eq.~(5) is the \emph{renormalized}
(long-wavelength) superfluid stiffness rather than the microscopic bare lattice
value.  Quantum Monte-Carlo simulations of the two-dimensional Bose gas show
that strong vortex-mediated phase fluctuations reduce the stiffness near the
BKT transition to only a fraction of its zero-temperature value \cite{Prokof_Svistunov_MC}.  Specifically, the renormalized stiffness satisfies
\begin{equation}
\rho_{s}^{\mathrm{eff}}(T_{\mathrm{BKT}}) 
= f(0)\,\rho_{s}^{(0)}(n),
\qquad 
f(0) \simeq 0.1\text{--}0.3,
\label{eq:renorm_stiffness}
\end{equation}
$f(0)\simeq 0.1$--$0.3$ is completely \emph{independent} of lattice geometry or microscopic interaction strength. Hard-core bosons on a triangular lattice, the XXZ model, and continuum two-dimensional bosons all flow to the same XY universality class at long wavelengths.  The same universal renormalization of the
stiffness applies to the vacancy-induced exciton superfluid. So that the characteristic BKT temperature is therefore,
\begin{equation}
T_{BKT}= \pi\sqrt{3}\, \frac{t}{k_{B}} n(1-n)f(0)    
\end{equation} 
implying transition temperatures on the order of 
$T_{\mathrm{BKT}}\!\sim\!35$--$100~\mathrm{K}$ for $f(0)\simeq 0.1$--$0.3$, at half filling $n=1/2$.
The predicted transition temperature above is fully consistent with other two-dimensional bosonic superfluids possessing comparable kinetic scales.  Bi-layer transition metal dichalcogenide excitons are predicted to undergo BKT transitions in the range $20$--$80~\mathrm{K}$~\cite{fogler2014high}. Quantum Monte Carlo studies of the two-dimensional Bose--Hubbard model find a BKT transition temperature of the order $T_{\mathrm{BKT}}\sim (0.1$--$0.3)\,t$ for hard-core bosons on a lattice~\cite{Capogrosso2008}. For $t=22~\mathrm{meV}$ this corresponds to $T_{\mathrm{BKT}}\sim 25$--$75~\mathrm{K}$, fully consistent with the range $35$--$100~\mathrm{K}$ obtained from Eq.~(10).

\paragraph{Conclusions---}We have shown that Te substitution in SL Nb$_3$Cl$_8$ introduces vacancies that eliminate the lowest-energy dark exciton and thereby enable exciton mobility on the SL breathing kagome lattice Nb$_3$Cl$_8$. The resulting fractional-filling regime is naturally described by the hard-core boson manifold of an effective XXZ model, whose transverse order corresponds to excitonic superfluidity. A continuum Landau--Ginzburg analysis demonstrates that the superfluid--normal transition is of BKT type, with a transition temperature set by the vacancy-induced hopping amplitude. These results identify vacancy engineering as a practical route to realizing a two-dimensional exciton superfluid in a single-layer transition-metal halide. Our findings open a pathway toward identifying experimental signatures of vacancy-driven exciton superfluidity in single-layer Nb$_3$Cl$_8$ derivatives, including vortex-mediated transport anomalies and characteristic optical responses near the BKT transition. Moreover, exciton superfluidity offers a charge-neutral transport channel that strongly suppresses impurity scattering and Joule heating, enabling long-lived spin and valley coherence and providing a promising route toward low-power excitronics and excitonic spintronics.

Beyond the superfluid phase below $T_{\mathrm{BKT}}$, the present platform naturally lends itself to exciton-based spintronics even in the normal state. Our BSE calculations yield a Frenkel exciton insulator state with energy $E_{\mathrm{EMI}}\simeq -0.14~\mathrm{eV}$ relative to the single-electron gap \cite{khan2024multiferroic}, i.e., a negative formation energy signaling that the spin–1 excitons remain thermodynamically stable up to and above room temperature, even when long-range phase coherence is destroyed by unbound vortices for $T>T_{\mathrm{BKT}}$. In this regime, the system realizes a dense, strongly correlated gas (or Mott liquid) of localized spin–1 excitons on the breathing–kagome lattice. This opens the door to a spin–1 excitonic spintronics platform that operates well above $T_{\mathrm{BKT}}$, where spin information is encoded in the occupation of the $m=0,\pm 1$ components of the triplet manifold and transported by incoherent exciton hopping.

\bibliographystyle{apsrev4-2}
\bibliography{Bibliography}

\end{document}